\documentclass[pra, aps, reprint]{revtex4-2}
\usepackage{bm}
\usepackage{graphicx}
\usepackage{siunitx}
\usepackage[final]{microtype}
\usepackage{amsmath, amsfonts, amssymb}
\usepackage{xcolor}
\usepackage{braket}

\newcommand\diff{\mathop{}\!d}
\DeclareSIUnit\wn{\raiseto{-1}\cm}

\begin{document}

\title{Cold collisions of rovibrationally excited D$_2$ molecules}

\author{James F. E. Croft}
\affiliation{The Dodd-Walls Centre for Photonic and Quantum Technologies, Dunedin 9016, New Zealand}
\affiliation{Department of Physics, University of Otago, Dunedin 9016, New Zealand}
\author{Pablo G. Jambrina}
\affiliation{Departamento de Qu\'{i}mica F\'{i}sica, Universidad de Salamanca, Salamanca 37008, Spain}
\author{F. Javier Aoiz}
\affiliation{Departamento de Qu\'{i}mica F\'{i}sica, Universidad Complutense, Madrid 28040, Spain}
\author{Hua Guo}
\affiliation{Department of Chemistry and Chemical Biology, University of New Mexico, Albuquerque, New Mexico 87131, USA}
\author{N. Balakrishnan}
\affiliation{Department of Chemistry and Biochemistry, University of Nevada,
Las Vegas, Nevada 89154, USA}


\begin{abstract}
The H$_2$+H$_2$ system has long been considered as a benchmark system for ro-vibrational energy transfer in bimolecular collisions. However, most studies thus far have focused on collisions involving  H$_2$ molecules in the ground vibrational level or in the first excited vibrational state.
While H$_2$+H$_2$/HD  collisions have received wide attention due to the important role they play in astrophysics, D$_2$+D$_2$ collisions have received much less attention. Recently, Zhou \emph{et al.} [Nat. Chem. {\bf 14} 658 (2022)]  examined stereodynamic aspects of rotational energy transfer in collisions of two aligned D$_2$ molecules prepared in the $v=2$ vibrational level and $j=2$ rotational level.
Here, we report  quantum calculations of rotational and vibrational energy transfer in collisions of two D$_2$ molecules prepared in vibrational levels up to $v=2$ and identify key resonance features that contribute to the angular distribution in the experimental results of Zhou \emph{et al.}\
The quantum scattering calculations were performed in full dimensionality and using the rigid-rotor approximation using a recently-developed highly-accurate six-dimensional potential energy surface for the H$_4$ system that allows descriptions of collisions involving highly vibrationally excited H$_2$ and its isotopologues.
\end{abstract}
\maketitle

\section{Introduction}\label{sec:intro}
Bimolecular collisions involving diatomic molecules are  topics of active investigation due to the continuing interest in cooling and trapping of molecules and their applications in sensing, precision spectroscopy, quantum information processing, and ultracold chemistry~\cite{Rev_Carr08,Rev_Krems08,Rev_Ye2014,Rev_Bala16,Rev_Bohn17,KRb+KRb_Science_2019,Toscano_Ultracold_Review_PCCP_2020,Yu_Ann_Rev_2021,Heazelwood_Ultracold_Review_2021,Kendrick_Li2Na_2021,KRb+KRB_NatChem_2021,liu.hu.ea:precision}.
Heteronuclear alkali-metal dimers due to their easily accessible optical transitions and open-shell molecules with their nearly-diagonal Franck-Condon factors permitting laser cooling and trapping are the molecules of choice in these experiments~\cite{Anderegg_CaF_2018,Blackmore_2018,CaF+CaF_PRL_2020,SrF_PRL_2021,CaFRb_PRL_2021}.
These systems are largely not amenable to full-quantum  scattering calculations due to their high density of states, chemically reactive nature, and possible non-adiabatic effects.
Thus, most theoretical studies of cold and ultracold diatom-diatom collisions involving these systems have resorted to approximate models or not explicitly considered geometries in which all four atoms are in close proximity~\cite{yang.huang.ea:statistical,croft.bohn.ea:unified}.
From a computational perspective, lighter molecular systems are preferable for explicit comparisons and benchmarking between theory and experiment. However,  molecules such as H$_2$ and D$_2$ that serve as benchmark systems for bimolecular collisions are not amenable to laser cooling or other molecule cooling methods that rely on properties such as a permanent dipole moment or magnetic moment.

In the last several years, Mukherjee, Zare, and coworkers have reported a series of experiments on rotational quenching of optically state prepared (and aligned) HD in collisions with unpolarized H$_2$, D$_2$ and He~\cite{2017_Science_Perreault,2018_NatChem_Perreault,SARP_HD-He,sarp_hed2,sarp_hed2_science,perreault.zhou.ea:coherent,zhou.perreault.ea:anisotropic}. More recently, they have extended this approach to D$_2$+He~\cite{sarp_hed2,sarp_hed2_science} and  D$_2$+D$_2$ collisions~\cite{zhou.perreault.ea:anisotropic} in which the D$_2$ molecule is prepared in the $v=2$ vibrational level and $j=2$ rotational level.
A scheme to optically prepare D$_2$ molecules in the $v=4$ vibrational level has also been proposed~\cite{perreault.zhou.ea:coherent}. For the first time, rotational transfer in D$_2(v=2,j=2)$+D$_2(v=2,j=2)$ collisions was reported in which state-preparation and the alignment of both D$_2$ molecules were controlled using the Stark-induced adiabatic Raman passage (SARP) technique~\cite{zhou.perreault.ea:anisotropic}. Additionally, co-expansion of the molecules (or the two colliding species) in the same molecular beam allows relative collision energies to be reduced to the vicinity of 1~K. This permits studies of cold molecular collisions involving H$_2$, HD and D$_2$ in the 1~K range where collisions are dominated by a few low-lying partial waves in the incident channel.
Further, the SARP method can align the molecular bond axis relative to the initial relative velocity vector allowing stereodynamic studies of molecular collisions in a regime where collision dynamics is dominated by isolated partial wave resonances.
For rotational transfer in D$_2(v=2,j=2)$+D$_2(v=2,j=2)$ collisions Zhou \emph{et al.}\ concluded that the angular distribution is primarily governed by a $l=2$ partial wave resonance in the incident channel at around 1~K.
However such conclusions, obtained by fitting outgoing partial wave scattering amplitudes to experimental angular distributions, may not tell the full story especially  due to the relatively broad collision energy distribution of the molecules in the beam.
In fact, as discussed in this manuscript and elsewhere, our results show that several resonances contribute in this regime, with  dominant contributions from an $l=4$ partial wave resonance~\cite{D2preprint}.

While H$_2$+H$_2$ and H$_2$+HD collisions have been the topics of numerous prior investigations, D$_2$+D$_2$ collisions have received much less attention.
A survey of previous studies of H$_2$+H$_2$ and H$_2$+HD collisions was provided in our recent work that reported a new full-dimensional potential energy surface (PES) for the H$_4$ system~\cite{zuo.croft.ea:full-dimensional}.  Early studies of H$_2$+H$_2$ collisions have used the rigid rotor (RR) approximation.
This includes the seminal work of Green~\cite{Green_H2H2_JCP_1975} that reported one of the earliest calculations of rotational transitions in H$_2$+H$_2$ collisions using the analytic interaction potential developed by Zarur and Rabitz~\cite{Zarur_H2H2_1974}.
This is followed by theoretical studies of Flower~\cite{Flower_H2H2_1998} and Flower and Roueff~\cite{Flower_Roueff_1998,Flower_Roueff_1999} using the interaction potential of Schwenke~\cite{Schwenke_H2H2_JCP_1988}.
Lee \emph{et al.}~\cite{lee.balakrishnan.ea:rotational} reported rotational  transitions using a more accurate four-dimensional (4D) PES of Diep and Johnson~\cite{DJ_H2H2_PES,DJ_H2H2_PES_Erratum}, referred to as the DJ PES.

The first full-dimensional quantum calculations of ro-vibrational transitions in H$_2$+H$_2$ collisions were reported by Pogrebnya and Clary~\cite{pogrebnya.clary:full-dimensional} within the coupled-states (CS) approximation using the six-dimensional (6D) potential energy surface of Bothroyd \emph{et al.}~\cite{boothroyd.dove.ea:accurate}, referred to as the BMKP PES.
However, the BMKP PES yielded  rate coefficients that are too large compared to experimental results owing to the presence of high-order anisotropic terms in the interaction potential and its less accurate description of the long-range interaction.
A restricted version of this PES, referred to as BMKPE PES~\cite{pogrebnya.clary:full-dimensional}, that excluded high-order anisotropic terms, yielded results in better agreement with experiments. Quantum wave packet calculations of ro-vibrational transitions within the CS approximation were also reported by Lin and Guo~\cite{ying-lin.guo:full-dimensional,lin.guo:full-dimensional*1,lin.guo:full-dimensional*2}  on the BMKP PES.
Gatti \emph{et al.}~\cite{gatti.otto.ea:rotational} and Panda \emph{et al.}~\cite{Panda_H2H2_JCP_2007} reported cross sections for rotational transitions  in \emph{para}-H$_2$+\emph{para}-H$_2$ and \emph{ortho}-H$_2$+\emph{para}-H$_2$  collisions using the multiconfiguration time-dependent Hartree (MCTDH) algorithm and the BMKP PES.
The BMKP PES was also used in full-dimensional coupled-channel (CC) calculations of ro-vibrational transitions in H$_2$+H$_2$ collisions~\cite{quemener.balakrishnan.ea:vibrational,quemener.balakrishnan:quantum}.
An interaction potential developed by Hinde~\cite{hinde:six-dimensional} has been used in a number of recent CC studies of rovibrational transfer in H$_2$+H$_2$ collisions by Qu{\'e}m{\'e}ner \emph{et. al}~\cite{Bala_H2H2-JCP_2011,santos.balakrishnan.ea:quantum,santos.balakrishnan.ea:vibration-vibration} that yielded good agreement with experiments for both pure rotational and ro-vibrational transitions.
A 4D interaction potential developed by Patkowski \emph{et al.}~\cite{patkowski.cencek.ea:potential} that employed a larger basis set than the Hinde PES also yielded results in good agreement with experiments for pure rotational transitions.
Extensive calculations of rotational transitions involving collisions of highly rotationally excited H$_2$ molecules were reported by Wan \emph{et al.}~\cite{Wan_2018}  employing the 4D PES of Patkowski \emph{et al.} This potential was also used in recent studies of H$_2$+H$_2$ collisions by Hern\'andez \emph{et al.}~\cite{Hernandez_H2H2_2021}.
None of the H$_2$-H$_2$ PESs discussed above are constructed to study collisions of H$_2$ and its isotopologues with vibrational excitation beyond the $v=1$ vibrational level.
Here, we report rotational and vibrational transitions in collisions involving D$_2$ molecules for vibrational levels $v=0$--2 using a recently reported full-dimensional H$_4$ PES~\cite{zuo.croft.ea:full-dimensional}.
This  potential was chosen in this work because it covers a large configuration space and is constructed specifically for collisions between molecules in excited vibrational levels up to $v=10$.

Cross section measurements for elastic scattering in $n$-D$_2$+$n$-D$_2$ and $o$-D$_2$+$o$-D$_2$ collisions were first reported by Johnson, Grace, and Skofronick~\cite{johnson.grace.ea:total} for relative velocities in the range of 190-1000 m/s ($\sim$ 4--121~K).
Accompanying theoretical calculations by the same authors using a model potential developed by Schaefer and Meyer~\cite{Schaefer_Meyer_JCP_1979} revealed a three  peak structure arising from orbital resonances  $l=4$ near 170~m/s, $\l=5$  near 280~m/s, and $l =6$ near 400~m/s.
While these resonance peaks were not resolved in the experiments the minimum between the $l=4$ and $l=5$ resonances was reproduced by the experimental data.
We are not aware of any subsequent theoretical studies that reproduced these resonance features or provided more rigorous comparisons with experiments.
To the best of our knowledge the only theoretical work on D$_2$+D$_2$ inelastic collisions corresponds to recent work of Montero and P\'er\'ez-R\'ios~\cite{Montero_D2D2_JCP_2014} within a rigid rotor formalism using the PES of Diep and Johnson~\cite{DJ_H2H2_PES,DJ_H2H2_PES_Erratum} to evaluate bulk viscosity and rotational relaxation rates.

The paper is organized as follows.
In Sec.~\ref{sec:methods}, we provide a brief description of the scattering calculations.
In Sec.~\ref{sec:ground} we present results for elastic collisions between D$_2$ molecules in their ground vibrational state,
and compare with the experimental results of Johnson \emph{et al.}~\cite{johnson.grace.ea:total}.
In Sec.~\ref{sec:excited} we present rotational quenching cross sections for collisions between D$_2$ molecules in excited vibrational states,
focussing on D$_2$ in $v=2$ to compare with the experimental results of Zhou \emph{et al.}~\cite{zhou.perreault.ea:anisotropic}.
Finally we conclude with a summary of our findings in Sec.~\ref{sec:conclusion}.

\section{Methods}\label{sec:methods}

\subsection{Potential Energy Surface}
\begin{figure}[tb]
\centering
\includegraphics[width=\columnwidth]{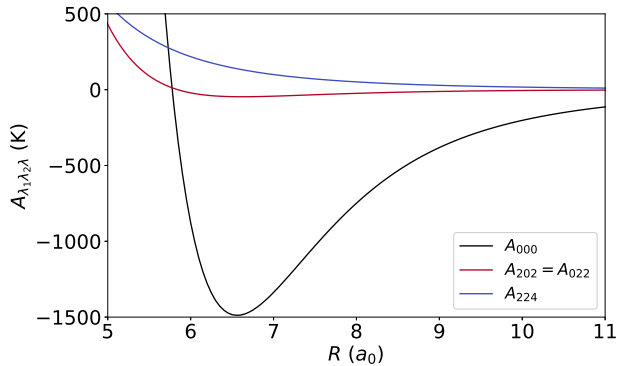}
\caption{Dominant terms in the spherical harmonic expansion of the angular dependence of the interaction potential as a function
of the distance between the two diatoms.}\label{fig:schematic}
\end{figure}

Computations were performed using the  full-dimensional H$_4$ PES of Zuo \emph{et al.}~\cite{zuo.croft.ea:full-dimensional}.
Fig.~\ref{fig:schematic} shows the dominant terms in the potential expansion with the D$_2$
distances fixed at their vibrational averaged values of $r=$\SI{1.435}{\bohr}
for the ground ro-vibrational state.
The angular dependence of the potential was expanded as

\begin{equation}
  U(\vec{r_1}, \vec{r_2}, \vec{R}) = \sum_\lambda A_\lambda(r_1, r_2, R)Y_\lambda(\hat{r}_1, \hat{r}_2, \hat{R}),
\end{equation}
with
\begin{align}
  Y_\lambda(\hat{r}_1, \hat{r}_2, \hat{R}) = \sum_m &  \braket{\lambda_1 m_1 \lambda_2 m_2| \lambda_{12} m_{12}} Y_{\lambda_1 m_1}(\hat{r}_1) \nonumber \\
  \times& Y_{\lambda_2 m_2}(\hat{r}_2) Y^*_{\lambda_{12} m_{12}}(\hat{R}),
\end{align}
where $\lambda \equiv \lambda_1\lambda_2\lambda_{12}$, $m \equiv m_1 m_2 m_{12}$,
$\vec{r_1}(r_1, \hat{r}_1)$ and $\vec{r_2}(r_2, \hat{r}_2)$ denote the vector
connecting the two Ds in each diatom while $\vec{R}(R, \hat{R})$ denotes the vector
joining the centers of mass of the two molecules.
For the scattering calculations it was found that $\lambda$ up to six was
sufficient to converge the cross sections for all transitions considered here.

\subsection{Scattering calculations}
Scattering calculations were performed in full-dimensionality using a modified
version of the TwoBC code~\cite{krems}.
The methodology for the collision between two indistinguishable molecules is
well established and has been outlined in detail
elsewhere~\cite{quemener.balakrishnan.ea:vibrational,
quemener.balakrishnan:quantum,santos.balakrishnan.ea:quantum,arthurs.dalgarno:theory,takayanagi:production,green:rotational,alexander.depristo:symmetry},
and been applied to other similar
systems involving various isotopologues of hydrogen~\cite{quemener.balakrishnan:quantum,
santos.balakrishnan.ea:vibration-vibration,balakrishnan.croft.ea:rotational,
croft.balakrishnan.ea:unraveling,croft.balakrishnan:controlling,
jambrina.croft.ea:stereodynamical,
zuo.croft.ea:full-dimensional}.
Here we briefly review the methodology in order to define notation.
The scattering calculations are performed within the
time-independent close-coupling formalism yielding the usual asymptotic
$S$ matrix~\cite{arthurs.dalgarno:theory}.
For convenience, we label each asymptotic channel by the
combined molecular state (CMS) $\alpha \equiv v_1 j_1 v_2 j_2$,
where $v$ and $j$ are the vibrational and rotational quantum numbers respectively
and the subscript refers to each molecule.
Since the molecules are indistinguishable we follow the ``well ordered'' states
classification, and include states where $v_1>v_2$ or when $v_1=v_2$,
$j_1 \ge j_2$~\cite{takayanagi:production,green:rotational,alexander.depristo:symmetry}.

The integral cross section for state-to-state rovibrationally inelastic
scattering for a given exchange permutation, $\epsilon_P$, and collision energy $E$, is given by,
\begin{align}
  \sigma^{\epsilon_P}_{\alpha \to \alpha'} &= \frac{\pi(1+\delta_{v_1v_2}\delta_{j_1j_2})(1+\delta_{v'_1v'_2}\delta_{j'_1j'_2})}{(2j_1+1)(2j_2+1)k_\alpha^2}\\ \nonumber  &\times \sum_{J,j_{12},j'_{12},l,l'}(2J+1) |T^{J \epsilon_P}_{\alpha lj_{12},\alpha'l'j'_{12}}|^2,
\end{align}
where $k^2=2 \mu E/\hbar^2$, $\mu$ is the reduced mass of the molecule-molecule system, $T^J = 1-S^J$, $l$ is the orbital angular momentum,
$J$ the total angular momentum ({\bf J} = {\bf l} + {\bf j$_{12}$}),
and {\bf j$_{12}$} = {\bf j}$_1$ + {\bf j}$_2$.
There are a number of conventions for the cross sections for collisions
between indistinguishable molecules that differ by factors of
2~\cite{monchick.schaefer:theoretical,green:rotational,huo.green:quantum,avdeenkov.bohn:ultracold}.
The difference has been discussed by a number of
authors~\cite{takayanagi:production,takayanagi:vibrational,rabitz.lam:rotational,
danby.flower.ea:rotationally,lee.balakrishnan.ea:rotational,tscherbul.suleimanov.ea:magnetic},
and all conventions yield the same physically observable rate.
We adopt here the convention of Green~\cite{quemener.balakrishnan:quantum}.
The corresponding energy dependent rate coefficient is
\begin{align}
\label{eq:non_thermal_rate}
  k^{\epsilon_P}_{\alpha \to \alpha'} = &\frac{\pi\hbar}{\mu k_\alpha(2j_1+1)(2j_2+1)}\\ \nonumber \times&\sum_{J,j_{12},j'_{12},l,l'}(2J+1) |T^{J \epsilon_P}_{\alpha lj_{12},\alpha'l'j'_{12}}|^2.
\end{align}

In general, experiments cannot select the nuclear spin state for the colliding
molecules and in which case the state-to-state cross section (rate) is given
by a statistically weighted sum of the exchange-permutation symmetrized cross
sections (rates):
\begin{align}\label{eqn:weights}
  \sigma_{\alpha \to \alpha'} &= W^+\sigma_{\alpha \to \alpha'}^+ + W^-\sigma_{\alpha \to \alpha'}^-, \\
  k_{\alpha \to \alpha'} &= W^+ k_{\alpha \to \alpha'}^+ + W^- k_{\alpha \to \alpha'}^-.
\end{align}
The D$_2$ molecule consists of two spin 1 D atoms, as such for collisions between
molecules in even rotational levels $W^+=21/36$ and $W^-=15/36$,
while for collisions between molecules in odd rotational levels $W^+=6/9$ and
$W^-=3/9$~\cite{johnson.grace.ea:total}.

Finally the experimental observable rate of molecules produced in $v_1'j_1'$
from $v_1 j_1$ per unit time is\begin{equation}
  \frac{\diff n_{v_1j_1 \to v_1'j_1'}}{\diff t} = \sum_{v_2' j'_2} \sum_{v_2j_2 } (1 + \delta_{v'_1v'_2}\delta_{j'_1j'_2}) k_{\alpha \to \alpha'}n_{v_1j_1 } n_{v_2j_2 },
\end{equation}
where the extra delta function accounts for the fact that each collision event
produces two molecules and $n_a$ refers to the density of molecules in state
$a$~\cite{stoof.koelman.ea:spin-exchange,burke-jr:theoretical}.

In  the scattering calculations the CC equations were propagated from 3 to \SI{103}{\bohr} with a
radial step size of \SI{0.05}{\bohr} using a log-derivative
method~\cite{manolopoulos:improved}.
The number of points in the radial coordinate for each diatom for the discrete
variable representation was 24;
the number of points in the angular coordinate $\theta$ between $\vec{R}$ and
$\vec{r}$ for each diatom for the Gauss-Legendre quadrature was 14;
the number of points in the dihedral angle between $\theta_1$ and $\theta_2$
for the Chebyshev quadrature was 8.
The rotational constant of D$_2$ is around \SI{41}{\kelvin}, as such the
D$_2$($j=4$) rotation level is around \SI{584}{\kelvin} above the $j=2$ level while the depth
of the interaction potential between two D$_2$ molecules is only around
\SI{30}{\kelvin}.
The basis set contained rotational levels up to $j=4$ within each vibrational manifold.
Scattering calculations were performed for each parity, exchange permutation
symmetry and total angular momentum quantum number up to and including 16.

\begin{figure}[tb]
\centering
\includegraphics[width=\columnwidth]{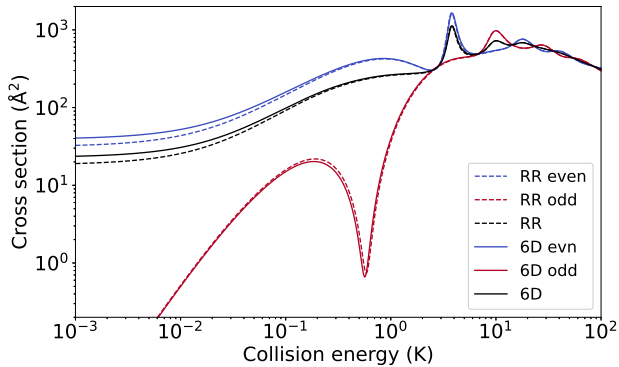}
\caption{
  Elastic cross section as a function of the collision energy for \emph{ortho}-D$_2$ + \emph{ortho}-D$_2$ collisions.
  Both the total cross section as well as the contribution from each exchange
  permutation symmetry are shown.
}\label{fig:oo}
\end{figure}
\begin{figure}[tb]
\centering
\includegraphics[width=\columnwidth]{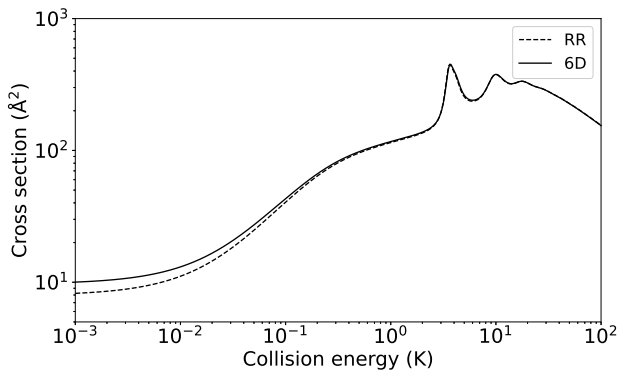}
\caption{
  Elastic cross section as a function of the collision energy for \emph{ortho}-D$_2$ + \emph{para}-D$_2$ collisions.
}\label{fig:op}
\end{figure}
\begin{figure}[tb]
\centering
\includegraphics[width=\columnwidth]{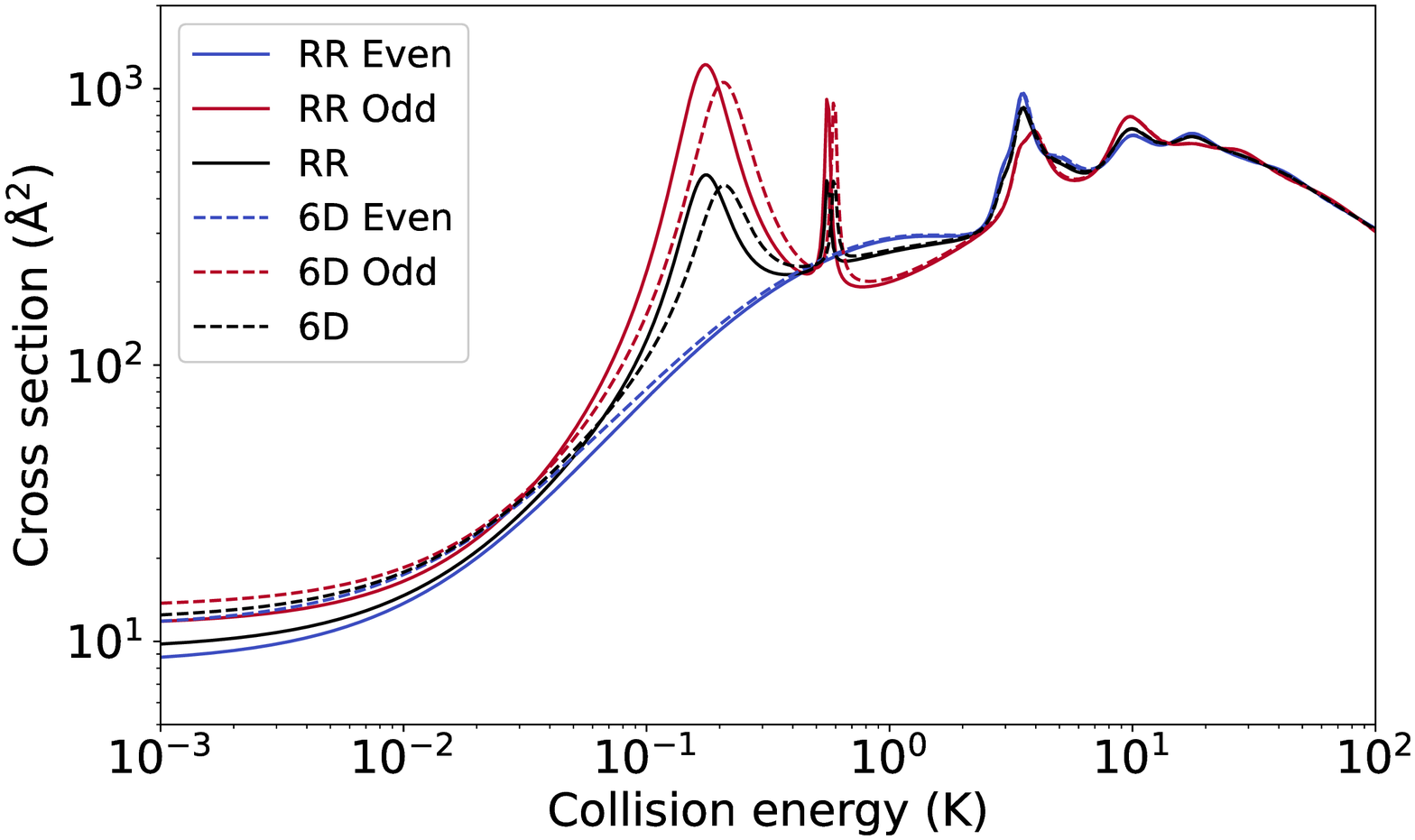}
\caption{
  Elastic cross section as a function of the collision energy for \emph{para}-D$_2$ + \emph{para}-D$_2$ collisions.
  Both the total cross section as well as the contribution from each exchange
  permutation symmetry are shown.
}\label{fig:pp}
\end{figure}

\begin{figure}[tb]
\centering
\includegraphics[width=\columnwidth]{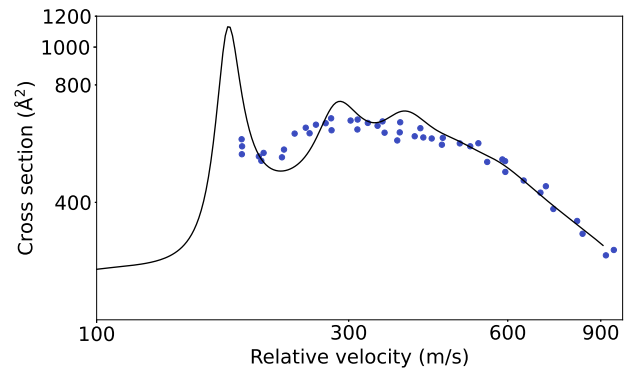}
\caption{
  Elastic cross section as a function of the relative collision velocity for
  \emph{ortho}-D$_2$ + \emph{ortho}-D$_2$ collisions.
  The blue dots are experimental results taken from Ref.~\cite{johnson.grace.ea:total}
  which have been scaled to those from the present calculations.
}\label{fig:oo_johnson}
\end{figure}

\section{Results}
\subsection{Elastic collisions between molecules in their ground vibrational state }\label{sec:ground}
By convention D$_2$ molecules in $j=0$ are called \emph{ortho}-deuterium (\emph{ortho}-D$_2$) while
D$_2$ molecules in $j=1$ are called \emph{para}-deuterium (\emph{para}-D$_2$)~\cite{johnson.grace.ea:total}.
Figs.~\ref{fig:oo}, \ref{fig:op}, and \ref{fig:pp} show elastic scattering
cross sections for \emph{ortho}-\emph{ortho} collisions (Fig.~\ref{fig:oo}),
\emph{ortho}-\emph{para} collisions (Fig.~\ref{fig:op}), and \emph{para}-\emph{para} collisions (Fig.~\ref{fig:pp}).
Collisions between \emph{ortho}-\emph{ortho} and \emph{para}-\emph{para} correspond to indistinguishable
molecules and Figs.~\ref{fig:oo} and \ref{fig:pp} show both the statistically weighted
total cross section as well as the separate even and odd exchange permutation cross sections.
Collisions between \emph{ortho} and \emph{para}-D$_2$ molecules are distinguishable since they
have different nuclear spins and Fig.~\ref{fig:op} shows just the total cross section.
Results are shown for the full 6D calculations as well as for rigid rotor
calculations with the D$_2$ internuclear distance fixed at the vibrational
averaged value of $r=$\SI{1.435}{\bohr} for the ground ro-vibrational state.
It is seen that except at low energies below $\sim$ \SI{1}{\kelvin} there are
no significant differences between the full 6D and the rigid rotor
calculations.

Figure~\ref{fig:oo_johnson} compares the present results for
\emph{ortho}-\emph{ortho} collisions with the experimental data
of Johnson \emph{et al.}~\cite{johnson.grace.ea:total} scaled to the present
theoretical cross sections.
Note that the experimental results are relative cross sections although
in Ref.~\cite{johnson.grace.ea:total} Johnson \emph{et al.} presented their
results scaled to calculations on several 1-D potentials.
The present calculations are in a general good agreement with those shown in Ref.~\cite{johnson.grace.ea:total}.

\begin{figure}[tb]
\centering
\includegraphics[width=\columnwidth]{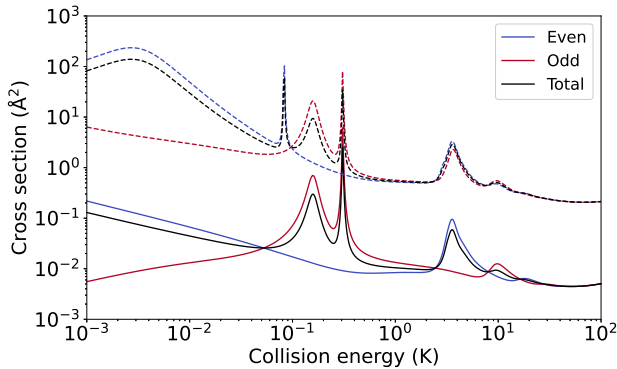}
\caption{%
  Rotational quenching cross sections as a function of the collision energy for D$_2$($v=0$, $j=2$) + D$_2$($v=0$, $j=2$) collisions.
  Showing both the total cross section as well as the contribution from each exchange permutation symmetry.
  The dashed lines are for the 0202 $\to$ 0002 while the solid lines are for the 0202 $\to$ 0000 transition.
}\label{fig:22}
\end{figure}

Figure~\ref{fig:22} shows the rotational quenching cross sections for collisions between
D$_2$($v=0$, $j=2$) molecules.
It is seen that even at the lowest energies shown (\SI{1}{\milli\kelvin}),
for the 0202 $\to$ 0002 collisions (using the CMS notation), the Wigner threshold regime ($\sigma \sim E^{-\frac{1}{2}}$ for $s$-wave collisions)
has not been reached~\cite{bethe:theory,wigner:on}.
This is generally the case when a near threshold resonance~\cite{simbotin.cote:jost} is present.
In this case the resonance appears as a broad peak at \SI{3}{\milli\kelvin}, and it is associated with $l=1$.
This resonance is not present for the 0202 $\to$ 0000 transitions, for which the Wigner threshold regime is reached at \SI{3}{\milli\kelvin}.

\subsection{Collisions between molecules in excited vibrational states.}\label{sec:excited}
\begin{figure}[tb]
\centering
\includegraphics[width=\columnwidth]{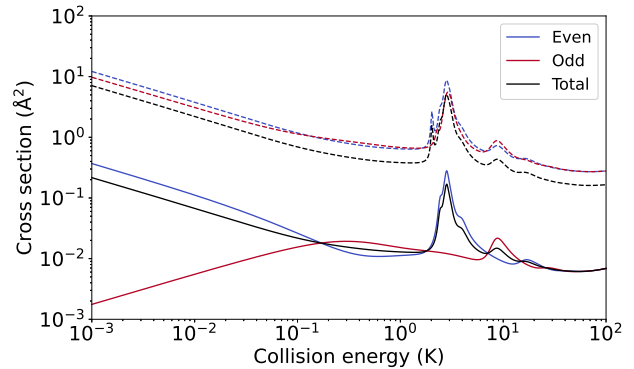}
\caption{
  Rotational quenching cross sections as a function of the collision energy for D$_2$($v=1$, $j=2$) + D$_2$($v=1$, $j=2$) collisions.
  Showing both the total cross section as well as the contribution from each exchange
  permutation symmetry.
  The dashed lines are for the 1212 $\to$ 1012 while the solid lines are for the 1212 $\to$ 1010 transition.
}\label{fig:1212}
\end{figure}

\begin{figure}[tb]
\centering
\includegraphics[width=\columnwidth]{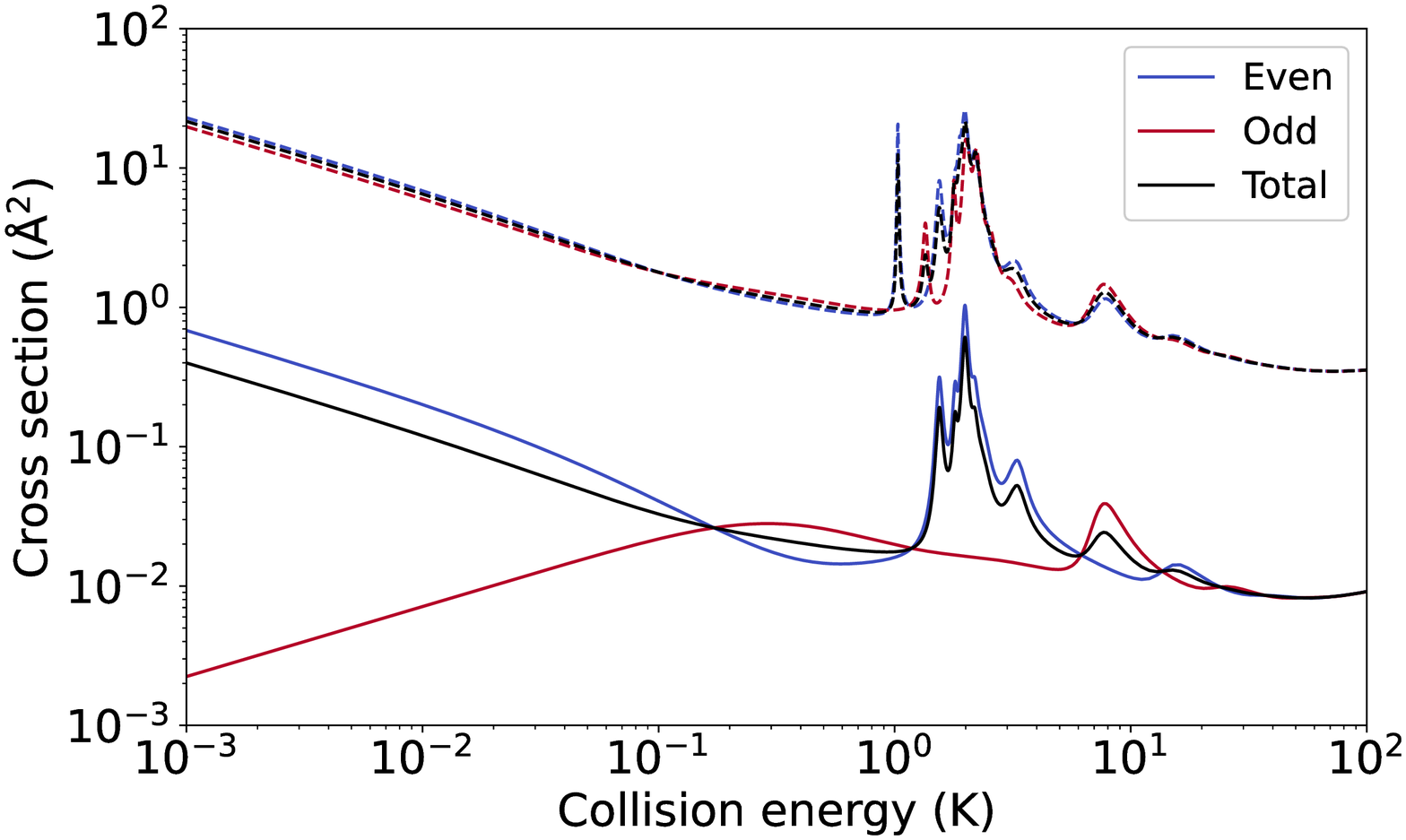}
\includegraphics[width=\columnwidth]{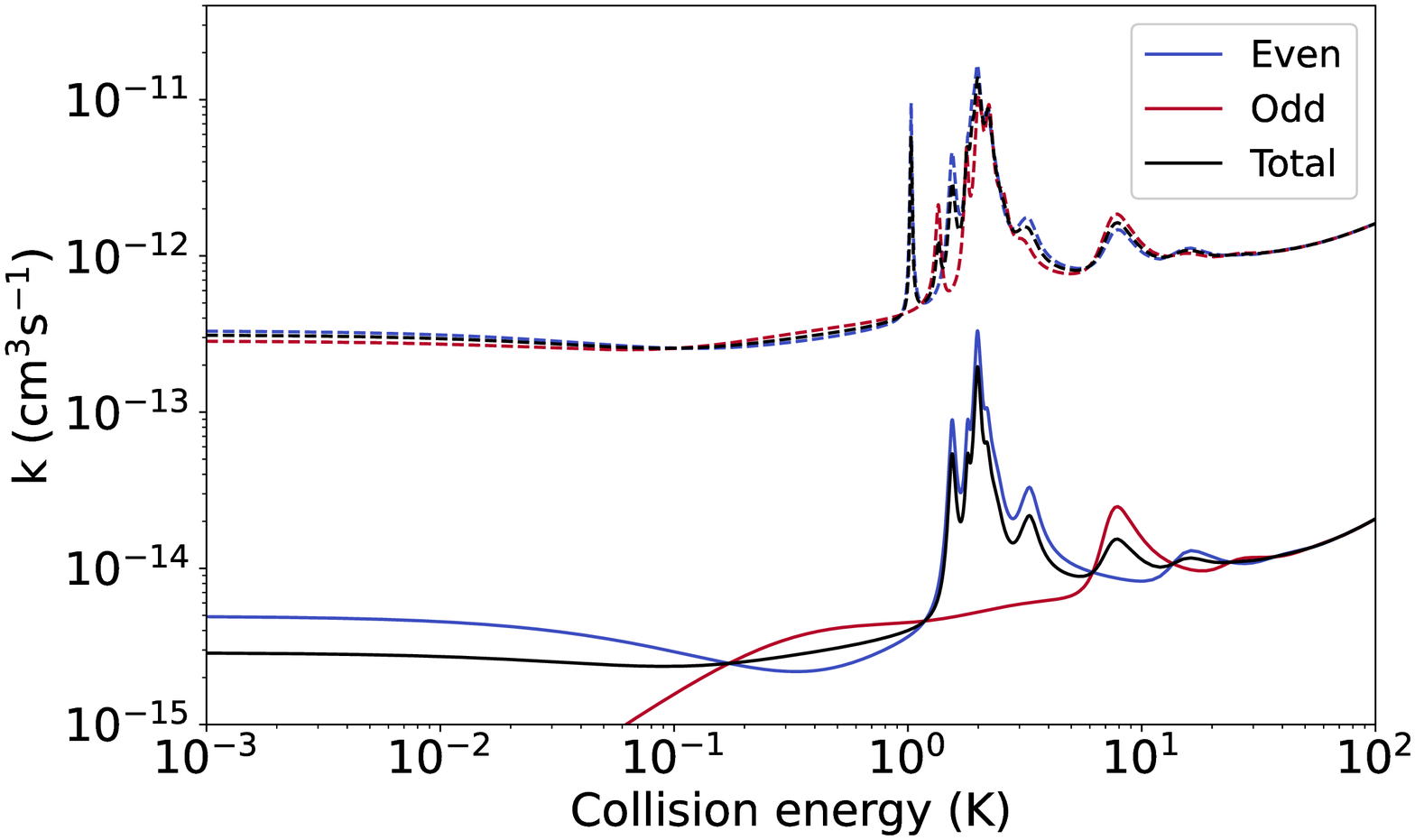}
\caption{%
  Rotational quenching cross sections (upper panel) and rates (lower panel)
  as a function of the collision energy for D$_2$($v=2$, $j=2$) + D$_2$($v=2$, $j=2$) collisions.
  Showing both the total cross section as well as the contribution from each exchange
  permutation symmetry.
  The dashed lines are for the 2222 $\to$ 2022 while the solid lines are for the 2222 $\to$ 2020 transition.
}\label{fig:2222}
\end{figure}

It is now possible using the SARP method to
achieve complete population transfer of molecules from the ground vibrational
level of D$_2$ to  excited vibrational
levels~\cite{zhou.perreault.ea:anisotropic,perreault.zhou.ea:coherent,mukherjee.zare:stark-induced}.
Since SARP is a two photon process the change in vibrational level is
accompanied by a $\Delta j =2$ rotational excitation.
As such here we examine collisions between two D$_2$ molecules in the same
vibrational excited state and both in $j=2$.
Figures~\ref{fig:1212} and~\ref{fig:2222} show the rotational quenching cross sections
for collisions between D$_2$ molecules in the first and second vibrational
excited states.
For collisions below \SI{100}{\kelvin} our results show that it was not necessary to
include basis functions corresponding to energetically closed vibrational levels.

Both Figs.~\ref{fig:1212} and~\ref{fig:2222} show the same general features,
however  significant differences are seen compared to the $v=0$ cross section shown in Fig.~\ref{fig:22}.
Unlike the $v=0$ case, for both $v=1$ and $v=2$, the Wigner threshold regime
is already reached by around \SI{0.1}{\kelvin}.
This appears to be due to the absence of a near-threshold resonance for the vibrationally excited collisions, that appear for $v=0$, $l=1$.
Such resonances are very sensitive to the channel potential and off-diagonal coupling matrix elements which vary with the initial ro-vibrational level.

\begin{figure}[tb]
\centering
\includegraphics[width=\columnwidth]{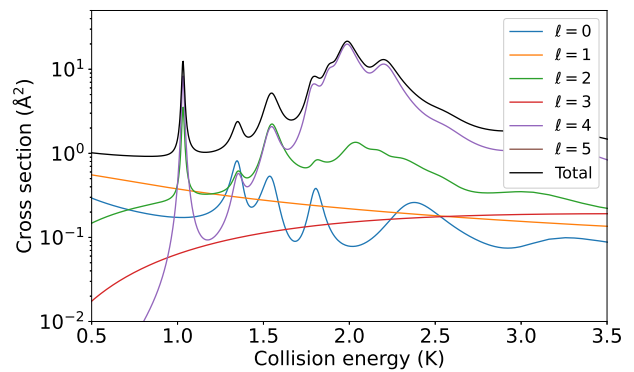}
\caption{%
$l$ resolved partial integral cross section for the 2222 $\to$ 2022 transition
as a function of the collision energy for D$_2$($v=2$, $j=2$) + D$_2$($v=2$, $j=2$) collisions.
}\label{fig:2222_lj}
\end{figure}

\begin{figure}[tb]
\centering
\includegraphics[width=\columnwidth]{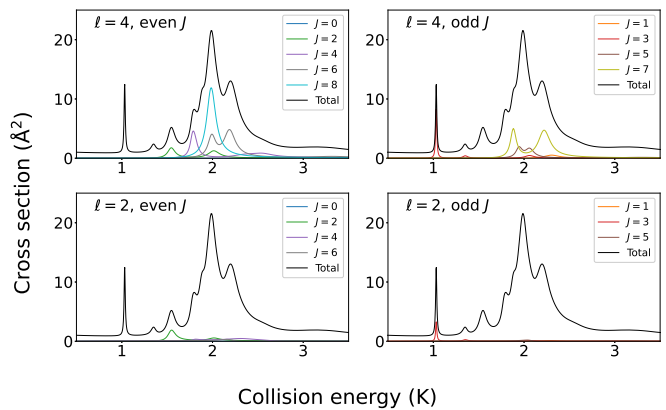}
\caption{%
$l$ and $J$ resolved partial integral cross sections for the 2222 $\to$ 2022 transition
as a function of the collision energy for D$_2$($v=2$, $j=2$) + D$_2$($v=2$, $j=2$) collisions.
The top two panels show the contribution from $l=4$ for even $J$ (left panel) and odd $J$ (right panel)
while the bottom two panels show the contribution from $l=2$ for even $J$ (left panel) and odd $J$ (right panel).
}\label{fig:2222_lj_panel}
\end{figure}

Figure~\ref{fig:2222} corresponds to the collisions experimentally studied by
Zhou \emph{et al.}~\cite{zhou.perreault.ea:anisotropic}.
They examined collisions in the 1--\SI{3}{\kelvin} range and
measured the angular distribution of scattered D$_2$($v'=2$, $j'=0$) molecules
(which are produced by both transitions shown in Fig.~\ref{fig:2222}) due to
collisions between aligned D$_2$ molecules.
It can be seen that the cross sections for the double rotational
quenching process are much smaller, and to a reasonable approximation
can be ignored in the analysis of the experimental data.
Analyzing the angular distribution they attributed the observed scattering
behavior to a $d$-wave ($l=2$) shape resonance~\cite{zhou.perreault.ea:anisotropic}.
In order to gain insight into the nature of the resonances seen in Fig.~\ref{fig:2222},
Fig.~\ref{fig:2222_lj} shows the $l$ resolved partial cross section, $\sigma^l(E)$.
The dominant contribution to the total cross section over the 1--\SI{3}{\kelvin}
is seen to be from $l=4$, consistent with the elastic cross sections reported by Johnson
\emph{et al.}~\cite{johnson.grace.ea:total},
as well as our previous analysis~\cite{D2preprint}.
The cluster of resonances at around \SI{2}{\kelvin} was present also for $v=0$ and $v=1$ collisions,
although in the case of $v=2$ it shows a dense structure which is associated with different $J$, for $l=4$.
Fig.~\ref{fig:2222_lj_panel} shows the contribution of different total angular momentum $J$ to the dominant
two partial waves ($l=2, 4$) seen in Fig.~\ref{fig:2222_lj}.
It is seen that the various peaks of the broad $l=4$ feature are due to contributions
from different total angular momentum $J$, while the narrow resonance at around \SI{1}{\kelvin} is due to
$J=3$ alone with contributions from $l=2$ and 4 and is absent in the double de-excitation.

While it is possible to switch off  the \SI{1}{K} resonance by a suitable preparation
of the internuclear axis of D$_2$ molecules, this is unlikely for the \SI{2}{K} group of resonances,
which have contributions from different values of $J$.
However, it is still possible to modulate the strength of the resonance peaks,
and it  would be  possible to split the overall resonance
peak, as demonstrated for H+HF collisions~\cite{jambrina.gonzalez-sanchez.ea:unveiling}.

\begin{figure}[tb]
\centering
\includegraphics[width=\columnwidth]{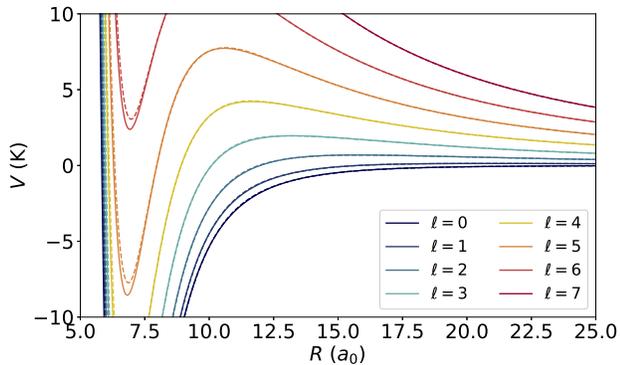}
\caption{%
  Isotropic part of the D$_2$ + D$_2$ potential + centrifugal barriers (solid lines),
  effective potential for $v_1=2, j_1=2, v_2=2, j_2=2, j_{12}=0, J=0$ (dashed lines).
}\label{fig:vl}
\end{figure}

In order to gain insight into the nature of the resonances seen in Figs.~\ref{fig:2222}
and~\ref{fig:2222_lj} we analyzed the isotropic part of the potential,
evaluated with D$_2$ distances kept at their vibrationally averaged values of
\SI{1.575}{\bohr} (for $v=2$, $j=2$),
\begin{equation}\label{eq:iso}
  V_l(R) = V_\mathrm{iso} + \frac{\hbar^2}{2\mu}\frac{l(l+1)}{R^2},
\end{equation}
where $V_\mathrm{iso} = A_{000}/{(4\pi)}^\frac{3}{2}$.
We also analyzed the effective potentials corresponding to different incoming
partial waves $l$,
\begin{align}\label{eq:eff}
V^J(R)=&\epsilon_{v_1j_1v_2j_2} \\  \nonumber
+&U^{J}_{v_1j_1v_2j_2 l j_{12}, v'_1 j' _1 v'_2 j'_2 l' j'_{12}}(R) + \frac{l(l+1)\hbar^2}{2\mu R^2}.
\end{align}
The first term is the energy of the CMS obtained by adding the asymptotic
rovibrational energies of the two D$_2$ molecules, the second term is the
potential energy matrix in the channel basis, and the third term is the
centrifugal potential for the orbital angular momentum $l$.
Fig.~\ref{fig:vl} shows both the radial potentials with the solid line showing the
isotropic potential (Eq.~\ref{eq:iso}) while the dashed line shows the
effective potential for $v_1=2, j_1=2, v_2=2, j_2=2, j_{12}=0, J=0$ (Eq.~\ref{eq:eff}).
The height of the centrifugal barriers are 0.13, 0.69, 1.96, 4.21, and
\SI{7.76}{\kelvin} for $l=1$ to 5 for the isotropic potential
(for the effective potential the barrier heights are almost
identical---0.13, 0.69, 1.96, 4.26, and \SI{7.77}{\kelvin}).
This suggests that in the  1--\SI{3}{\kelvin} region it is $l=3$ and 4
partial waves which will be the dominant contributions to the experimental cross section.
It is clearly seen that the radial potentials differ little suggesting
that using the isotropic part of the potential captures the main physics of
the problem.
The depth of the radial potentials increases with vibrational quantum number.
This explains why the low energy $l=1$ resonance seen for the 0202 collisions
(see Fig.~\ref{fig:22}) is not seen for the 1212 (see Fig.~\ref{fig:1212}) or
2222 (see Fig.~\ref{fig:2222}) cases since the resonance has been shifted to
below threshold.

\begin{figure}[tb]
\centering
\includegraphics[width=\columnwidth]{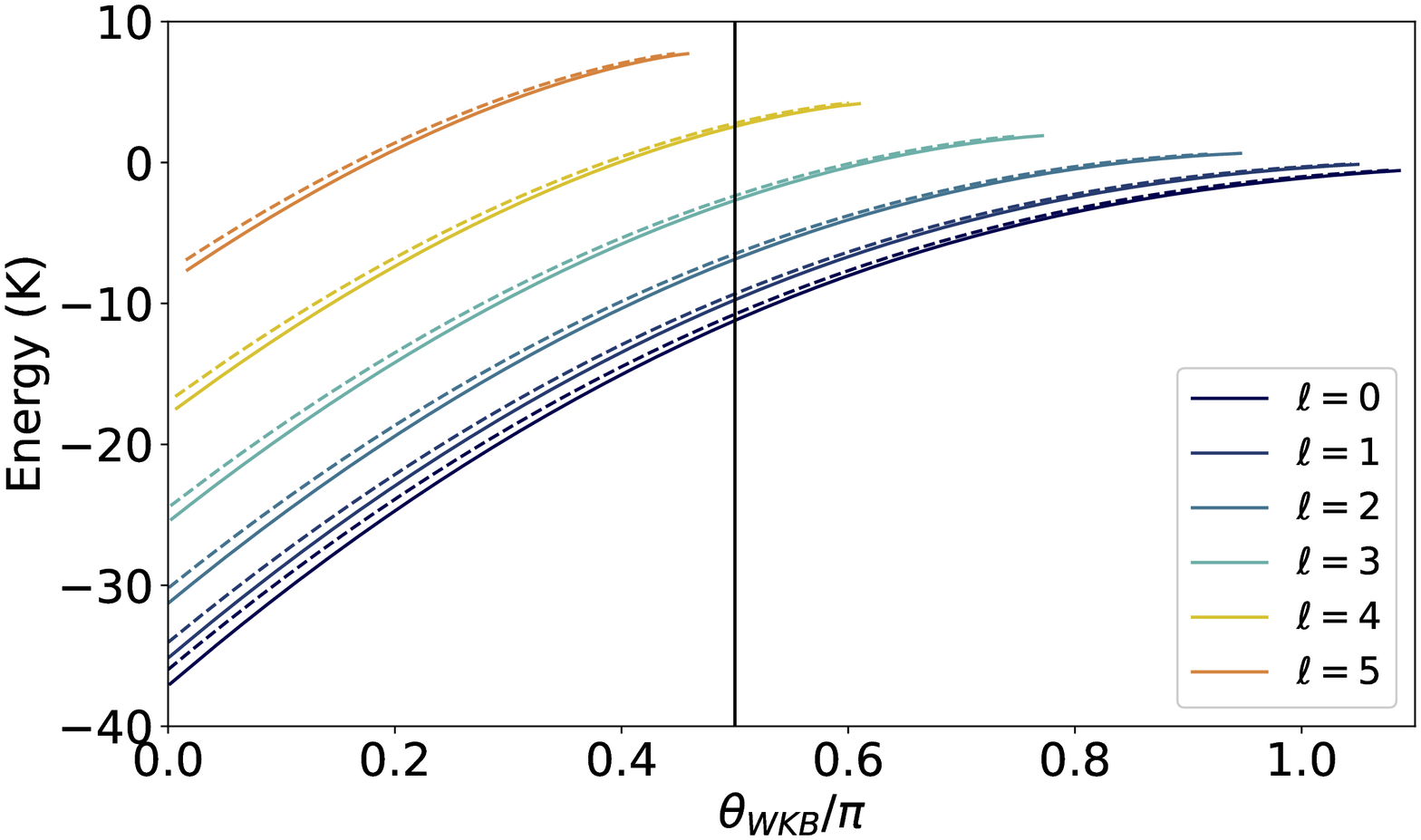}
\caption{%
  The WKB phase for the radial potentials shown in Fig.~\ref{fig:vl}.
  The vertical line shows the quantization condition for the lowest energy
  state.
}\label{fig:phase}
\end{figure}

To further investigate the nature of the resonances seen in Fig~\ref{fig:2222}
and~\ref{fig:2222_lj} we performed a semiclassical WKB analysis to find the
approximate locations of bound states for the system.
The WKB phase was evaluated from the integral
\begin{equation}
  \theta_\mathrm{WKB} = \int \diff R \sqrt{2\mu (V(R) - E)/\hbar^2},
\end{equation}
where the integration is carried out between the classical turning points at the energy $E$ for the radial potentials
shown in Fig.~\ref{fig:vl}.
The corresponding quantization condition is
$\theta_\mathrm{WKB} = \pi(n + \frac{1}{2})$, for the $n^\mathrm{th}$ bound
state of the potential.
Figure~\ref{fig:phase} shows the WKB phase as a function of the energy for
each of the radial potentials shown in Fig.~\ref{fig:vl}, along with a vertical
line identifying the quantization condition.
It can be seen that WKB predicts at most a single bound state in each channel
with a bound state for $l=0,1,2,3$ while an above threshold quasi-bound state
at around \SI{2}{\kelvin} for $l=4$, which manifests itself as a shape
resonance in the scattering experiments, in excellent agreement with the
full 6D scattering results shown in Figs.~\ref{fig:2222} and~\ref{fig:2222_lj}.
For $l=5$ there is no quasi-bound state with the WKB phase not reaching
$\pi/2$ below the height of the centrifugal barrier.

This analysis shows that for these types of systems it is possible to
predict the relevant partial waves that will contribute to the cross section
in the \SI{1}{\kelvin} range using simply the isotropic part of the potential
combined with a WKB analysis.
This suggests that using such an approach to identify the relevant partial waves
to inform an experimental analysis would be fruitful.


\section{Conclusion}\label{sec:conclusion}
We have presented full-dimensional scattering calculations of rotational and vibrational transitions in collisions between two D$_2$ molecules prepared in vibrational levels $v=0$--2 and rotational levels $j=0$--2.
Computations are performed using a recently reported full-dimensional \emph{ab initio} potential energy surface for the H$_4$ system designed to treat collisions of highly vibrationally excited H$_2$ and its  isotopologues.
Our calculations show that resonance features in low energy collisions of two D$_2$ molecules below 5~K arise primarily from a $l=4$ partial wave resonance, consistent with the prediction of Johnson \emph{et al.}~\cite{johnson.grace.ea:total} for elastic collisions using an isotropic potential.
For collisions between two D$_2$ molecules in the $v=2$ vibrational state and $j=2$ rotational state,
cross sections for double rotational relaxation is found to be about two orders of magnitude smaller than that of relaxation of only one molecule.

\section*{Acknowledgments}
This work was supported in part by NSF grant No. PHY-2110227  (N.B.) and ARO MURI
grant No. W911NF-19-1-0283 (N.B., H.G.). P.G.J.  gratefully acknowledges grant
PID2020-113147GA-I00 funded by MCIN/AEI/10.13039/, and F.J.A. acknowledges   funding
by the Spanish Ministry of Science and Innovation (Grant No. PGC2018-096444-B-I00 and PID2021-122839NB-I00).
J.F.E.C. gratefully acknowledges support from the Dodd-Walls Centre for Photonic and Quantum Technologies.

\bibliographystyle{naturemag}
\bibliography{refs}

\end{document}